\documentclass[twocolumn,10pt]{IEEEtran}

\usepackage[acronym]{glossaries}
\usepackage{amssymb}
\usepackage{amsmath}
\usepackage{graphicx}
\usepackage{booktabs}

\newacronym{awgn}{AWGN}{additive white Gaussian noise}
\newacronym{bs}{BS}{base station}
\newacronym{cnn}{CNN}{convolutional neural network}
\newacronym{gnn}{GNN}{graph neural network}
\newacronym{csi}{CSI}{channel state information}
\newacronym{dl}{DL}{deep learning}
\newacronym{fdd}{FDD}{frequency division duplex}
\newacronym{los}{LoS}{line-of-sight}
\newacronym{mimo}{MIMO}{multiple-input multiple-output}
\newacronym{ml}{ML}{machine learning}
\newacronym{nlos}{NLoS}{non-line-of-sight}
\newacronym{nn}{NN}{neural network}
\newacronym{snr}{SNR}{signal-to-noise ratio}
\newacronym{ue}{UE}{user equipment}
\newacronym{upa}{UPA}{uniform planar array}
\newacronym{cscg}{CSCG}{circularly symmetric complex Gaussian}
\newacronym{mmwave}{mmWave}{millimeter wave}
\newacronym{lidar}{LIDAR}{light detection and ranging}
\newacronym{aoa}{AoA}{angle of arrival}
\newacronym{aod}{AoD}{angle of departure}
\newacronym{tof}{ToF}{time of flight}
\newacronym{rf}{RF}{radio frequency}
\newacronym{dft}{DFT}{discrete Fourier transform}
\newacronym{gps}{GPS}{global positioning system}
\newacronym{sbr}{SBR}{shooting and bouncing rays}
\newacronym{mlp}{MLP}{multilayer perceptron}
\newacronym{svd}{SVD}{singular value decomposition}
\newacronym{rt}{RT}{ray tracing}
\newacronym{v2i}{V2I}{vehicle-to-infrastructure}
\newacronym{bn}{BN}{batch normalization}
\newacronym{relu}{ReLU}{rectified linear unit}
\newacronym{lrelu}{LReLU}{leaky rectified linear unit}
\newacronym{mae}{MAE}{mean absolute error}
\newacronym{flop}{FLOP}{floating point operation}

\begin{document}

\title{Overhead-Free Blockage Detection and Precoding Through Physics-Based Graph Neural Networks: LIDAR Data Meets Ray Tracing}

\author{Matteo Nerini,~\IEEEmembership{Student Member,~IEEE},
        Bruno Clerckx,~\IEEEmembership{Fellow,~IEEE}

\thanks{M. Nerini is with the Department of Electrical and Electronic Engineering, Imperial College London, London SW7 2AZ, U.K. (e-mail: m.nerini20@imperial.ac.uk).}
\thanks{B. Clerckx is with the Department of Electrical and Electronic Engineering, Imperial College London, London SW7 2AZ, U.K. and with Silicon Austria Labs (SAL), Graz A-8010, Austria (e-mail: b.clerckx@imperial.ac.uk).}}

\maketitle

\begin{abstract}
In this letter, we address blockage detection and precoder design for \gls{mimo} links, without communication overhead required.
Blockage detection is achieved by classifying \gls{lidar} data through a physics-based \gls{gnn}.
For precoder design, a preliminary channel estimate is obtained by running ray tracing on a 3D surface obtained from \gls{lidar} data.
This estimate is successively refined and the precoder is designed accordingly.
Numerical simulations show that blockage detection is successful with 95\% accuracy.
Our digital precoding achieves 90\% of the capacity and analog precoding outperforms previous works exploiting \gls{lidar} for precoder design.
\end{abstract}

\glsresetall


\section{Introduction}

Accurate precoding is critical to unlocking the full benefits of \gls{mimo} systems.
In \gls{fdd} systems, the precoder is typically designed through a closed-loop protocol.
The \gls{csi} is estimated at the receiver on the basis of pilot signals and fed back to the transmitter.
With \gls{csi} knowledge, the transmitter is finally able to select the proper precoder.
Another closed-loop protocol for precoder design is beam training, a process in which the candidate precoders are tested before selecting the optimal one. 
Beam training is common in \gls{mmwave} systems, where the receiver cannot observe the full channel matrix because of the subspace sampling problem \cite{hur13}.
However, these closed-loop approaches may cause a significant communication overhead due to the high antenna number in massive \gls{mimo} links \cite{hea16}.

In recent literature, side information has been exploited to reduce the overhead in the precoder design process.
To select the optimal precoder in \gls{mmwave}, sub-6GHz band signaling has been used in \cite{gon17,ali18}.
In \cite{va18}, the authors propose to select candidate beams as a function of the vehicle’s location by inverting the fingerprinting localization process.
Finally, vision-aided beam prediction has been proposed in \cite{alr20,cha21}.

\Gls{lidar} is a sensor commonly used in autonomous driving.
It employs a laser to scan the surrounding area and obtain a 3D point cloud.
\gls{lidar} is typically used for obstacle detection but can be also exploited to improve wireless communications.
In \cite{kla19}, \cite{dia19}, a \gls{cnn} has been used to identify the optimal beams given the \gls{lidar} data.
The authors used \gls{dl} strategies to design a reduced set of precoders to be tested during the beam training procedure.
As a result, the communication overhead caused by beam training can be significantly reduced.
In \cite{mas21}, the same problem has been solved through a federated learning approach, in which the connected vehicles collaborate to train the shared \gls{dl} model. 
In \cite{zec22}, the authors integrate knowledge distillation techniques, non-local attention schemes, and curriculum training to further improve the beam selection from \gls{lidar} data.

In this study, we address the problems of blockage detection and precoder design by completely removing the communication overhead.
This is realized by exploiting side information available at the transmitter, derived from \gls{lidar} and \gls{gps} data.
For blockage detection, we classify the \gls{lidar} point cloud through a physics-based \gls{gnn}, a \gls{dl} architecture never applied in this context.
For precoder design, we propose a novel joint use of \gls{lidar} data and ray tracing never investigated before.
First, we reconstruct a 3D surface representing the propagation environment from the \gls{lidar} point cloud.
Second, we obtain a preliminary channel estimate by running a ray tracing simulation on top of this surface.
Third, this preliminary channel estimate is refined through a \gls{cnn}, and the precoder is designed accordingly.
The joint use of \gls{lidar} data and ray tracing allows us to design the precoder with a physics-based \gls{dl} strategy, differently from state-of-the-art works relying on purely data-based approaches \cite{kla19}-\cite{zec22}.
By creating a \textit{digital twin} of the channel, our strategy is agnostic about the type of precoding considered, unlike related works selecting the best analog precoder within a predefined codebook \cite{kla19}-\cite{zec22}.
\section{System Model}
\label{sec:system-model}

Let us consider a point-to-point \gls{mimo} link between an $N_T$ antenna transmitter and an $N_R$ antenna receiver, in a \gls{v2i} scenario.
The transmitter is a vehicular \gls{ue} equipped with a \gls{lidar} and a localization system, such as \gls{gps}, while the receiver is a stationary \gls{bs}.
%
We denote the transmitted signal as $\sqrt{E_s}\mathbf{x}\in\mathbb{C}^{N_{T}\times1}$, where $E_s$ is the energy normalization factor and $\mathbf{x}$ is the precoded signal subject to the constraint $\left\|\mathbf{x}\right\|_2=1$.
The covariance matrix of the precoded signal $\mathbf{x}$ is denoted as the transmit covariance matrix and writes as $\mathbf{Q}=\mathbb{E}\left[\mathbf{x}\mathbf{x}^H\right]$.
Denoting the received signal as $\mathbf{y}\in\mathbb{C}^{N_{R}\times1}$, we have $\mathbf{y}=\sqrt{E_s}\mathbf{H}\mathbf{x}+\mathbf{n}$ where $\mathbf{H}\in\mathbb{C}^{N_{R}\times N_{T}}$ is the channel matrix, and $\mathbf{n}\sim\mathcal{CN}\left(\mathbf{0},\sigma_{n}^{2}\mathbf{I}\right)$ is the \gls{awgn}.
We write the channel matrix as $\mathbf{H}=\Lambda\widetilde{\mathbf{H}}$, where the scalar $\Lambda$ contains the path loss and shadowing, while $\widetilde{\mathbf{H}}$ accounts for the small-scale fading such that $\mathbb{E}\left[\|\widetilde{\mathbf{H}}\|_F^2\right]=N_RN_T$.
The achievable rate $R$ is given by
\begin{equation}
R=\log_2\det\left(\mathbf{I}+\rho\widetilde{\mathbf{H}}\mathbf{Q}\widetilde{\mathbf{H}}^H\right),\label{eq:rate}
\end{equation}
where $\rho=E_s\Lambda^2/\sigma_n^2$ is the \gls{snr}.

Given a channel estimate $\widehat{\mathbf{H}}$ at the \gls{ue}, two possible precoding strategies are considered: multi-stream digital precoding and analog precoding with a single \gls{rf} chain.
In the case of multi-stream digital precoding, the transmit covariance matrix maximizing \eqref{eq:rate} is given by multiple eigenmode transmission with water-filling power allocation, which is capacity achieving in the case of perfect \gls{csi}.
In the case of analog precoding, the transmit covariance matrix is $\mathbf{Q}=\mathbf{w}\mathbf{w}^H$, where $\mathbf{w}\in\mathbb{C}^{N_{T}\times1}$ is the analog precoder subject to the constraint $\mathbf{w}=\left(1/\sqrt{N_T}\right)\left[e^{j\varphi_1},e^{j\varphi_2},\dots,e^{j\varphi_{N_T}}\right]^T$, with $\varphi_i\in\left[0,2\pi\right)$ denoting the $i$-th phase shift.
We assume that the analog precoder $\mathbf{w}$ is chosen from a \gls{dft} codebook $\mathcal{W}$, obtained by adapting the codebook proposed in \cite{lov03} to $N_X\times N_Y$ \glspl{upa}.
This codebook is constructed as follows, depending on the number of quantization bits $B$ used for each phase shift $\varphi_i$.
We introduce the matrix $\mathbf{W_X}$ as the first $N_X$ rows of the $2^{B\left(N_X-1\right)}\times2^{B\left(N_X-1\right)}$ unitary \gls{dft} matrix.
In a similar way, $\mathbf{W_Y}$ is defined as a function of $N_Y$.
Thus, the codewords in $\mathcal{W}$ are given by the columns of the matrix $\mathbf{W}=\mathbf{W_X}\otimes\mathbf{W_Y}$, where $\otimes$ denotes the Kronecker product.
Given a channel estimate $\widehat{\mathbf{H}}$, the precoder $\mathbf{w}^\star$ maximizing \eqref{eq:rate} writes as
\begin{equation}
\mathbf{w}^\star=\arg\max_{\mathbf{w}\in\mathcal{W}}\mathbf{w}^H\widehat{\mathbf{H}}^H\widehat{\mathbf{H}}\mathbf{w}.\label{eq:analog-precoder}
\end{equation}

Our objective is to acquire a channel estimate $\widehat{\mathbf{H}}$ at the transmitter by completely removing the communication overhead caused by traditional closed-loop protocols for \gls{csi} acquisition.
This is achieved by exploiting the information from the \gls{lidar} and the \gls{gps} sensors, which are available at each transmitter.

\section{LIDAR Data Processing}
\label{sec:lidar}

\begin{figure}[t]
    \centering
    \includegraphics[width=0.23\textwidth]{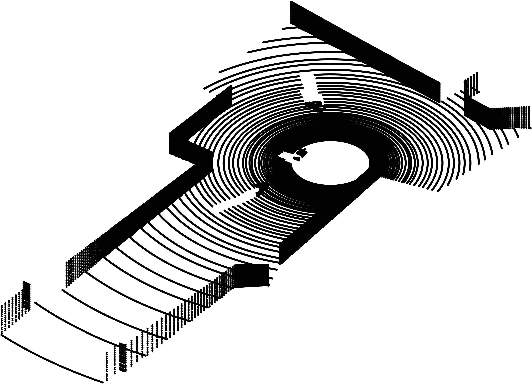}
    \includegraphics[width=0.23\textwidth]{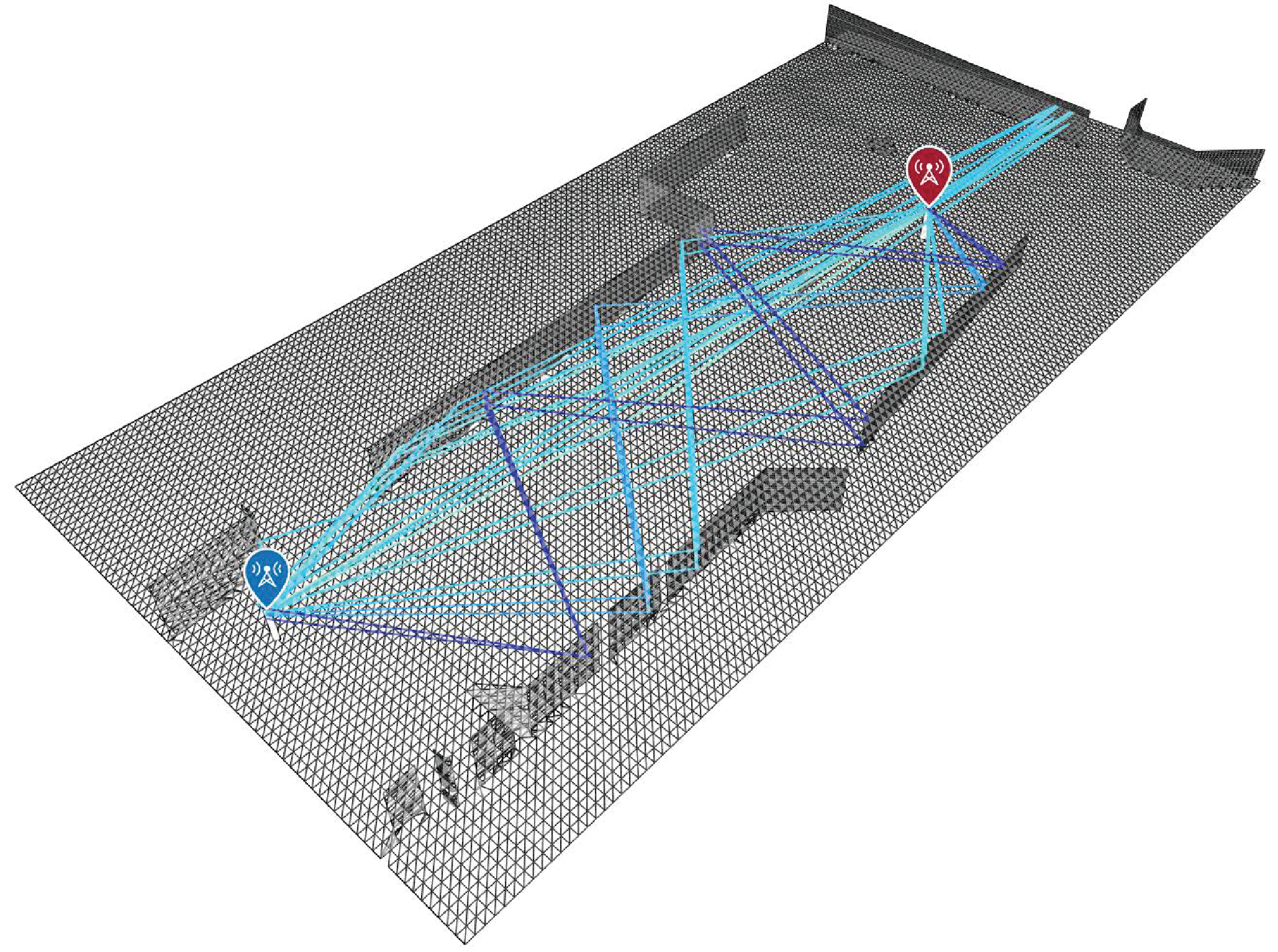}
    \caption{Example of LIDAR point cloud (left) and relative ray tracing simulation output (right).}
    \label{fig:lidar}
\end{figure}

As a result of the \gls{lidar} measurements, a point cloud $\mathcal{P}=\{\left(x_i,y_i,z_i\right)\}_{i=1}^{\left|\mathcal{P}\right|}$ is available at the transmitter.
In this study, we use this point cloud to reconstruct a 3D surface representing the physical environment in which the transmitter and receiver are located.
Based on the reconstructed 3D surface, a ray tracing simulation is locally run at the transmitter assuming that the location of the transmitting \gls{ue} and receiving \gls{bs} are known at the transmitter.
Finally, the resulting channel paths are combined to obtain a channel estimate at the transmitter, with no communication overhead.
Fig.~\ref{fig:lidar} shows an example of \gls{lidar} point cloud and the channel paths obtained by running ray tracing on the reconstructed 3D surface.

With the ray tracing simulation, we obtain $L_{RT}$ channel paths, each described by a vector of features.
Specifically, the $l$-th path is characterized by the complex gain $\alpha_l$, the azimuth \gls{aoa} $\theta_{R,l}$, the elevation \gls{aoa} $\phi_{R,l}$, the azimuth \gls{aod} $\theta_{T,l}$, and the elevation \gls{aod} $\phi_{T,l}$.
In addition, ray tracing outputs also the \gls{los} status of the $l$-th path as $s_{RT,l}\in\{0,1\}$.
We have $s_{RT,l}=1$ if the $l$-th path is the \gls{los} path between the \gls{ue} and the \gls{bs}, and $s_{RT,l}=0$ otherwise.
The array steering vector $\mathbf{a}\left(\theta,\phi\right)$ of \glspl{upa} with dimensions $N_{X}\times N_{Y}$ writes as
\begin{equation}
\mathbf{a}\left(\theta,\phi\right)=\frac{1}{\sqrt{N_XN_Y}}
\begin{bmatrix}
1\\
e^{j\Omega_X}\\
\vdots\\
e^{j\left(N_X-1\right)\Omega_X}\\
\end{bmatrix}
\otimes
\begin{bmatrix}
1\\
e^{j\Omega_Y}\\
\vdots\\
e^{j\left(N_Y-1\right)\Omega_Y}\\
\end{bmatrix},\label{eq:steering-vector}
\end{equation}
where $\Omega_X=kd_x\text{sin}\left(\theta\right)\text{cos}\left(\phi\right)$ and $\Omega_Y=kd_y\text{sin}\left(\theta\right)\text{sin}\left(\phi\right)$, with $k=2\pi/\lambda$ denoting the wave number.
In this study, we set the antenna spacing $d_x=d_y=\lambda/2$.
Given the $L_{RT}$ channel paths derived from the ray tracing simulation and the steering vector in \eqref{eq:steering-vector}, we compute the resulting \gls{mimo} channel according to the 3D \gls{mmwave} channel model \cite{hea16}.
The resulting narrowband channel matrix $\mathbf{H}_{RT}\in\mathbb{C}^{N_{R}\times N_{T}}$ writes as
\begin{equation}
\mathbf{H}_{RT}=\sqrt{N_RN_T}\sum_{l=1}^{L_{RT}}\alpha_l\mathbf{a}_R\left(\theta_{R,l},\phi_{R,l}\right)\mathbf{a}_T^H\left(\theta_{T,l},\phi_{T,l}\right),\label{eq:channel}
\end{equation}
which is regarded as the channel estimate derived from the ray tracing simulation.

\section{Blockage Detection}
\label{sec:blockage-detection}

\gls{mmwave} technology highly relies on the presence of the \gls{los} link to establish effective communication.
Thus, it becomes crucial to assess whether the channel is in \gls{los} or in \gls{nlos}.
We denote the channel status as $s\in\{0,1\}$, where $s=1$ means \gls{los} and $s=0$ means \gls{nlos}.
In this section, we propose three strategies to estimate the channel status $s$ given the \gls{lidar} data.

The first strategy uses the output of the ray tracing simulation carried out as described in Section~\ref{sec:lidar}.
Let us assume that the resulting $L_{RT}$ channel paths have been ordered such that their path gains $\vert\alpha_l\vert$ are decreasing.
We observe that among the channel paths, at most one is the \gls{los} path with $s_{RT,l}=1$.
In addition, if such a path is present, it must be the first path since longer trajectories result in lower gains.
Thus, the channel status estimate $\hat{s}_{RT}$ is given by $\hat{s}_{RT}=s_{RT,1}$.
We refer to this strategy as purely physics-based.

The second strategy is referred to as purely data-based since it employs a \gls{dl} architecture to process the \gls{lidar} point cloud.
The blockage detection problem is formalized as a point cloud binary classification problem and solved through a \gls{gnn}.
We adopt a \gls{gnn} since it is the only \gls{dl} architecture designed to effectively deal with data lying on manifolds, point clouds, and graphs.
Our \gls{gnn} processes the data through a three-stage scheme.
In the first stage, grouping stage, the \gls{lidar} point cloud $\mathcal{P}$ is transformed into an undirected graph $\mathcal{G}$ by connecting each point to its $k$-nearest neighbors.
In the second stage, neighborhood aggregation stage, each point aggregates information from its connected neighbors through the so-called message passing scheme.
We refer to the set of points connected to the $i$-th point in $\mathcal{G}$ as the neighborhood of $i$, which is denoted as $\mathcal{N}(i)$.
Furthermore, we define the position of the $i$-th point as $\mathbf{p}_i=[x_i,y_i,z_i]\in\mathbb{R}^3$.
In our implementation, we consider a two-hop message passing in which the hidden features of the $i$-th point in the $\ell$-th hop, with $\ell\in\{1,2\}$, is given by
\begin{equation}
\mathbf{h}^{(\ell)}_i=\sum_{j\in\mathcal{N}(i)\cup\{i\}}\textrm{MLP1}\left(\left[\mathbf{h}_j^{(\ell-1)},\mathbf{p}_j-\mathbf{p}_i\right]\right),
\end{equation}
where we set $\mathbf{h}_j^{(0)}=\mathbf{p}_j$.
$\textrm{MLP1}$ denotes a \gls{mlp} composed by four layers, all with 32 neurons and \gls{relu} activation.
After the message passing process, each point holds information about its two-hop neighborhood.
In the third stage, global aggregation stage, we apply a global graph readout function to aggregate all the node feature vectors $\mathbf{h}^{(2)}_i$, with $i=1,\ldots,\left|\mathcal{P}\right|$, into a unique graph feature vector $\mathbf{h}$.
To this end, we extract $\mathbf{h}=\max\{\mathbf{h}_1^{(2)},\ldots,\mathbf{h}_{\left|\mathcal{P}\right|}^{(2)}\}$, where the function $\max$ takes the element-wise maximum.
Finally, we apply a classifier to map the remaining features to one of the two status \gls{los} or \gls{nlos} as $\hat{s}_{GNN}=\textrm{MLP2}\left(\mathbf{h}\right)$.
$\textrm{MLP2}$ denotes a \gls{mlp} composed by two layers: the first with 16 neurons and \gls{relu} activation function; the second with one neuron and Sigmoid activation function.
The output of $\textrm{MLP2}$ $\hat{s}_{GNN}$ gives the channel status estimate.
To train the \gls{gnn}, we minimize a loss function given by the binary cross-entropy between $\hat{s}_{GNN}$ and the ground truth $s$.

The third strategy considers the integration of both physics-based and data-based approaches, referred to as physics-based \gls{dl}.
To this end, we slightly modify the \gls{gnn} used in our purely data-driven strategy.
After the global aggregation stage, the feature vector $\mathbf{h}$ and the physics-based status estimates $\hat{s}_{RT}$ are concatenated.
The resulting vector $\left[\hat{s}_{RT},\mathbf{h}\right]$ is fed in input to a new classifier $\textrm{MLP3}$ estimating the channel status as $\hat{s}_{PBGNN}=\textrm{MLP3}\left(\left[\hat{s}_{RT},\mathbf{h}\right]\right)$, where $\textrm{MLP3}$ denotes a \gls{mlp} with the same structure as $\textrm{MLP2}$.
Finally, $\hat{s}_{PBGNN}$ is the channel status estimate given by our physics-based \gls{gnn}.

\section{Overhead-Free Channel Estimation}
\label{sec:precoder}

We now consider the problem of estimating the channel without communication overhead.
Since our ultimate objective is to design the precoder, as discussed in Section~\ref{sec:system-model}, it is sufficient to acquire the estimate $\widehat{\mathbf{T}}=\widehat{\mathbf{H}}^H\widehat{\mathbf{H}}$.
We propose to compute $\widehat{\mathbf{T}}$ by refining the preliminary channel estimate derived from the ray tracing simulation.
To this end, we introduce the matrix $\mathbf{T}_{RT}=\mathbf{H}_{RT}^H\mathbf{H}_{RT}$ with eigenvalue decomposition $\mathbf{T}_{RT}=\mathbf{V}_{RT}^H\mathbf{\Lambda}_{RT}\mathbf{V}_{RT}$.

To obtain $\widehat{\mathbf{T}}$, we first estimate the blockage status of the channel through the physics-based \gls{gnn} strategy presented in Section~\ref{sec:blockage-detection}.
Then, $\widehat{\mathbf{T}}$ is computed differently for channels classified as \gls{los} and \gls{nlos}, since these two channel types are characterized by different properties.
In \gls{los} channels, the power is mainly captured by the \gls{los} path.
Thus, they are expected to be well reconstructed by ray tracing since the reflected paths, though inaccurate, have a minimal impact.
Conversely, in \gls{nlos} channels, inaccurate ray tracing might decrease the channel reconstruction quality.
For this reason, we set the estimate $\widehat{\mathbf{T}}=\widehat{\mathbf{T}}_{LoS}$ if $\hat{s}_{PBGNN}\geq1/2$ and $\widehat{\mathbf{T}}=\widehat{\mathbf{T}}_{NLoS}$ if $\hat{s}_{PBGNN}<1/2$.
In the following, we illustrate how $\widehat{\mathbf{T}}_{LoS}$ and $\widehat{\mathbf{T}}_{NLoS}$ are computed.

For channels classified as \gls{los}, we define $\mathbf{H}_{LoS}\in\mathbb{C}^{N_{R}\times N_{T}}$ as the rank-1 channel given by the \gls{los} path between \gls{ue} and \gls{bs}.
The channel $\mathbf{H}_{LoS}$ is computed by plugging only the \gls{los} path into \eqref{eq:channel}.
Given $\mathbf{H}_{LoS}$, the channel estimate in \gls{los} is refined by forcing the dominant eigenvector of $\mathbf{T}_{RT}$ to be the unique right eigenvector of $\mathbf{H}_{LoS}$, denoted as $\mathbf{v}_{LoS}\in\mathbb{C}^{N_{T}\times 1}$.
More precisely, we set $\widehat{\mathbf{T}}_{LoS}=\widehat{\mathbf{V}}_{LoS}^H\widehat{\mathbf{\Lambda}}_{LoS}\widehat{\mathbf{V}}_{LoS}$, where $\widehat{\mathbf{\Lambda}}_{LoS}=\mathbf{\Lambda}_{RT}$ and $\widehat{\mathbf{V}}_{LoS}=GS([\mathbf{v}_{LoS},[\mathbf{V}_{RT}]_{:,2:N_{T}}])$.
Here, $GS([\mathbf{v}_{LoS},[\mathbf{V}_{RT}]_{:,2:N_{T}}])$ denotes the Gram-Schmidt process applied to the columns of the concatenation $[\mathbf{v}_{LoS},[\mathbf{V}_{RT}]_{:,2:N_{T}}]$.

\begin{figure}[t]
    \centering
    \includegraphics[width=0.4\textwidth]{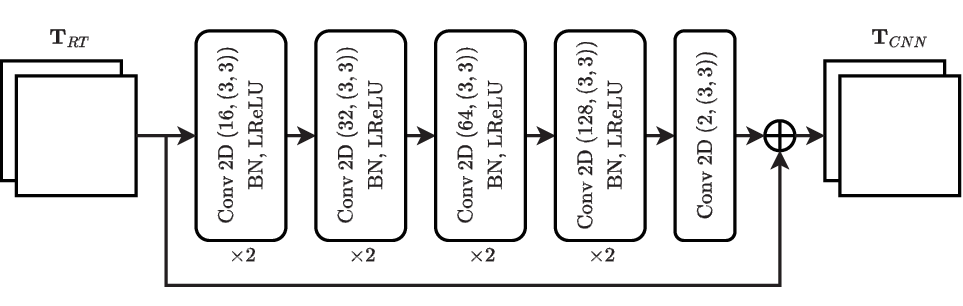}
    \caption{Proposed CNN model for channel refinement.}
    \label{fig:cnn}
\end{figure}

For channels classified as \gls{nlos}, the ray tracing estimate is refined through a \gls{cnn}, applying the principle of physics-based \gls{dl}.
Our \gls{cnn} receives in input the matrix $\mathbf{T}_{RT}$, organized as a two-channel $N_T\times N_T$ real tensor.
This input is processed by a series of convolutional layers, able to capture the spatial redundancy in $\mathbf{T}_{RT}$.
These layers are placed in parallel to a skip connection to avoid overfitting, as represented in Fig.~\ref{fig:cnn}.
The hidden layers are characterized by an increasing depth, spanning from $16$ to $128$ kernels.
Following each hidden layer, we place a \gls{bn} and a \gls{lrelu} activation.
The output layer has depth $2$ and linear activation, since it must approximate the difference $\mathbf{T}_{CNN}-\mathbf{T}_{RT}$, where $\mathbf{T}_{CNN}$ is the refined version of $\mathbf{T}_{RT}$ given by the \gls{cnn} output.
In all the layers, the kernel size is $3\times 3$, with unitary strides and zero-padding such that the output of each layer has the same size as the input.
The \gls{cnn} has been trained by minimizing a loss function given by the \gls{mae} between $\mathbf{T}_{CNN}$ and the ground truth $\mathbf{T}=\mathbf{H}^H\mathbf{H}$, which can be obtained with dedicated sampling campaigns in real-world scenarios.
However, we note that imposing such a Hermitian positive semi-definite constraint to the output of a \gls{cnn} is hard.
Thus, we add a post-processing operation to transform $\mathbf{T}_{CNN}$ to its nearest Hermitian positive semi-definite matrix.
The nearest Hermitian positive semi-definite matrix in the Frobenius norm to an arbitrary matrix $\mathbf{T}_{CNN}$ is given by $\widehat{\mathbf{T}}_{NLoS}=(\mathbf{S}+\mathbf{P})/2$, where $\mathbf{S}=(\mathbf{T}_{CNN}+\mathbf{T}_{CNN}^H)/2$ is the nearest Hermitian matrix to $\mathbf{T}_{CNN}$, and $\mathbf{P}$ is the Hermitian polar factor of $\mathbf{S}$.

Given $\widehat{\mathbf{T}}$, the digital precoder is computed according to multiple eigenmode transmission with water-filling power allocation, while the analog precoder is selected from the codebook according to \eqref{eq:analog-precoder}.
Differently from beam training, the closed-loop beam selection strategy widely adopted in \glspl{mmwave}, our strategy is overhead-free and enables both digital and analog precoding.


\section{Numerical Results}
\label{sec:results}


\subsection{Simulation Methodology}

\begin{figure}[t]
    \centering
    \includegraphics[width=0.23\textwidth]{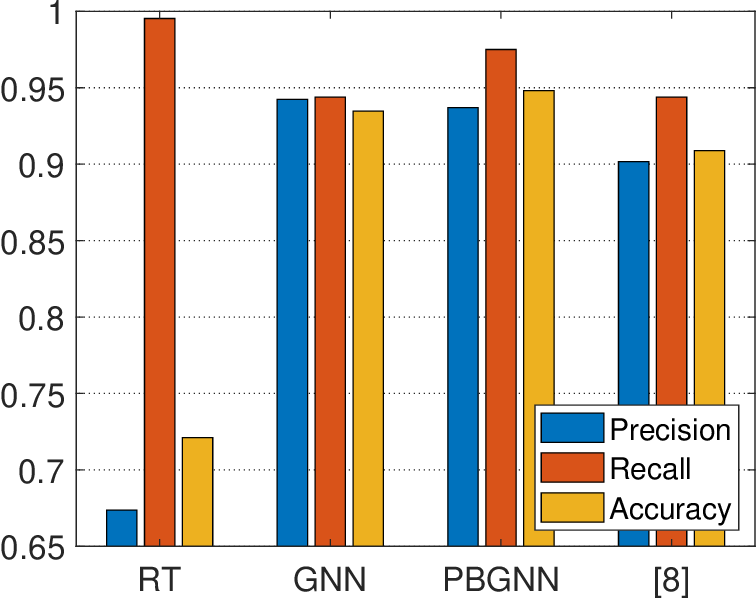}
    \caption{Performance of blockage detection solutions in terms of precision, recall, and accuracy.}
    \label{fig:block-det}
\end{figure}

To evaluate the performance of the proposed approaches, we employ the benchmark dataset Raymobtime s008 \cite{kla18}.
This dataset is composed of 11194 channel realizations paired with their respective \gls{lidar} and \gls{gps} data, among which 6482 are in \gls{los} and 4712 are in \gls{nlos} status.
We build the training, validation, and test set with $80\%$, $10\%$, and $10\%$ of the samples, respectively. 
The antennas at the \gls{bs} and at the \gls{ue} are $4\times4$ \glspl{upa}, yielding $N_T=N_R=16$.
Furthermore, the \gls{dft} codebook of analog precoders has been generated by setting $B=2$, resulting in a set of $4096$ precoders.
The value of $B$ has been selected to provide a good trade-off between resolution and the number of precoders.
To decrease the codebook cardinality, we prune this initial codebook by retaining only the useful precoders.
Specifically, we retain only the precoders that are optimal more than twice in the training set.
This procedure reduces the codebook cardinality from $4096$ to $218$ possible precoders.

To reconstruct the open 3D surface we adopted the mesh-growing algorithm proposed in \cite{dia11}.
This algorithm proved to be particularly fast and accurate on point clouds of open surfaces.
Given the 3D surface representing the environment geometry, the channel paths have been computed by employing the ray tracing propagation model available in Matlab.
The \gls{sbr} method has been used to calculate the channel paths with up to 6 path reflections.


The trainable parameters of the proposed \gls{dl} models have been tuned by minimizing the loss function on the training set.
Through cross-validation on the validation set, we have set the hyperparameters of the models characterizing their architectures, as described in Sections~\ref{sec:blockage-detection} and \ref{sec:precoder}.
Further hyperparameters have been set as follows.
For the \gls{gnn} employed for blockage detection, we set the neighborhood size in the graph $\mathcal{G}$ to $k=8$.
We train the \gls{gnn} using the Adam optimizer with learning rate $10^{-3}$ for 100 epochs, with batch size 64.
The \gls{cnn} employed to refine the ray tracing channel has been trained solely on the \gls{nlos} channels, using the Adam optimizer with learning rate $10^{-3}$ for 200 epochs, with batch size 200.

\subsection{Performance Evaluation}

We evaluate the performance of blockage detection in terms of precision, recall, and accuracy.
In Fig.~\ref{fig:block-det}, we assess the performance of our three strategies compared with the \gls{cnn} proposed in \cite{kla19}.
The physics-based strategy ``RT'' achieves an extremely high recall but a low precision.
This strategy rarely classifies a \gls{los} channel as \gls{nlos} (a false negative) since the \gls{lidar} 3D surface rarely contain non-existing obstacles between transmitter and receiver.
Conversely, it is likely to miss existing obstacles, classifying a \gls{nlos} channel as \gls{los} (a false positive).
The data-based strategy ``GNN'' outperforms the \gls{cnn} proposed in \cite{kla19}.
In the physics-based \gls{dl} strategy ``PBGNN'', the knowledge of $\hat{s}_{RT}$ produces a twofold effect.
On the one hand, $\hat{s}_{RT}$ helps the neural network to correctly classify \gls{los} channels, increasing the recall with respect to ``GNN''.
On the other hand, the precision is slightly decreased since $\hat{s}_{RT}$ produces a high number of false positives.
Overall, the accuracy shows that ``PBGNN'' outperforms ``GNN''.

\begin{table}[t]
\centering
\caption{Complexity of blockage detection solutions.}
\begin{tabular}{@{}cccc@{}}
\toprule
Model & \multicolumn{2}{c}{FLOPs} & \# Params.\\
      & Aggregation net. & Classifier net. &\\
\midrule
GNN   & $494\times10^3$ & $1.07\times10^3$ & $8.26\times10^3$\\
PBGNN & $494\times10^3$ & $1.11\times10^3$ & $8.27\times10^3$\\
\cite{kla19} & \multicolumn{2}{c}{$1028\times10^6$} & $148\times10^3$\\
\bottomrule
\end{tabular}
\label{tab:complexity-LoS}
\end{table}

In Tab.~\ref{tab:complexity-LoS}, we compare our strategies with \cite{kla19} in terms of \glspl{flop} and trainable parameters.
In \glspl{gnn}, the total number of \glspl{flop} depends on the considered point cloud since an aggregation network (implementing the two-hop message passing) is instantiated for every point in the point cloud.
For this reason, we separately provide the \glspl{flop} in the aggregation and classifier networks.
In the Raymobtime s008 dataset, the average number of points in \gls{lidar} point clouds is $E[\left|\mathcal{P}\right|]=815$, yielding approximately $403\times 10^6$ total \glspl{flop} for both ``GNN'' and ``PBGNN'', on average \cite{kla18}.
Thus, our blockage detection solution is attractive in terms of accuracy, computational complexity, and stored parameters.

\begin{figure}[t]
    \centering
    \includegraphics[width=0.23\textwidth]{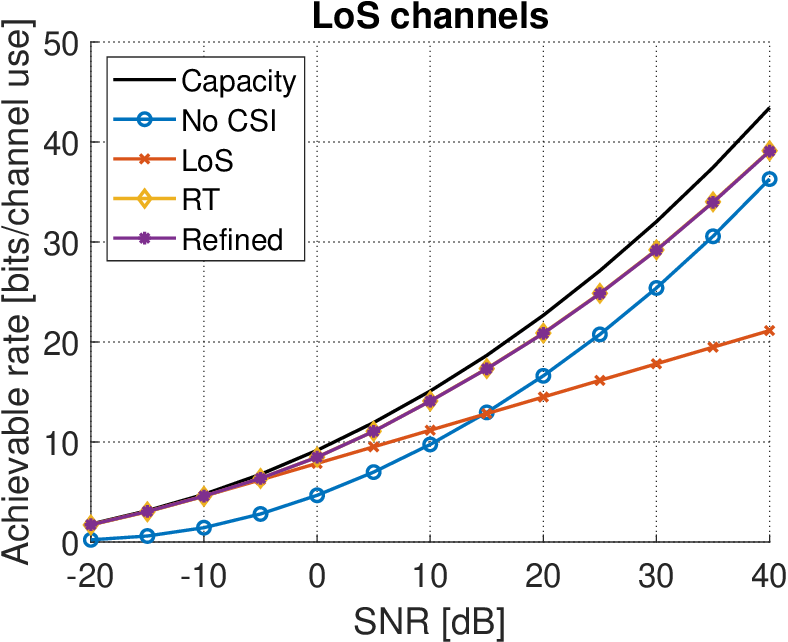}
    \includegraphics[width=0.23\textwidth]{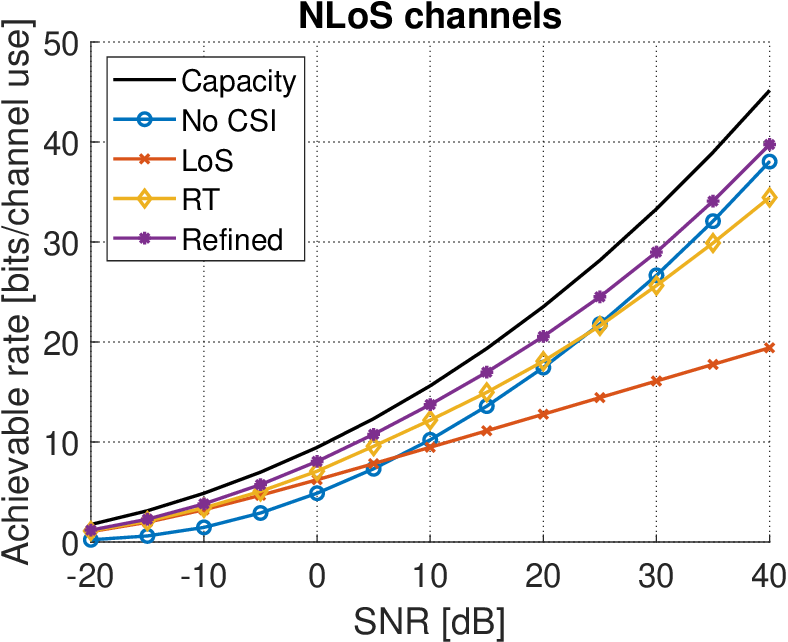}
    \caption{Achievable rate for LoS channels (left) and NLoS channels (right) obtained with digital precoding.}
    \label{fig:digital}
\end{figure}

The performance of digital precoding is evaluated considering the achievable rate $R$ as a metric.
In Fig.~\ref{fig:digital}, we report the performance of digital precoding based on our overhead-free channel estimation strategy ``Refined''.
We also report the channel capacity, achieved with perfect \gls{csi}, and the following baselines: \emph{i)} ``No CSI'': The transmit covariance matrix is $\mathbf{Q}=1/N_T\mathbf{I}$; \emph{ii)} ``LoS'': The precoding is based on the rank-1 \gls{los} channel $\mathbf{H}_{LoS}$; and \emph{iii)} ``RT'': The precoding is based on the ray tracing channel estimate $\mathbf{H}_{RT}$.
Fig.~\ref{fig:digital} shows that our channel refinement strategy outperforms the ``No CSI'' and ``LoS'' baselines.
The benefit versus the ``RT'' strategy is visible for \gls{nlos} channels, where the \gls{cnn} is used to refine the ray tracing channel estimate.
Our channel refinement strategy allows achieving approximately $90\%$ of the capacity without any communication overhead in both \gls{los} and \gls{nlos} channels.

\begin{figure}[t]
    \centering
    \includegraphics[width=0.23\textwidth]{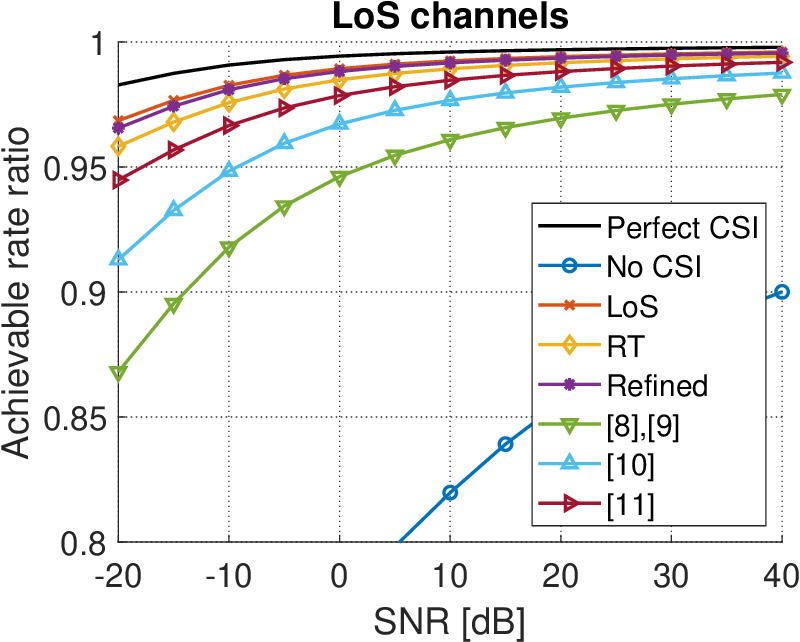}
    \includegraphics[width=0.23\textwidth]{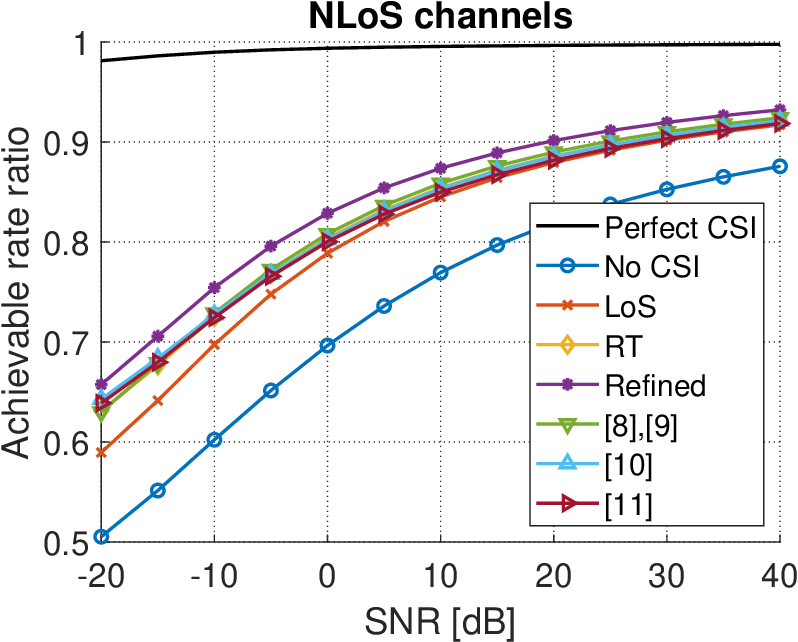}
    \caption{Achievable rate ratio for LoS channels (left) and NLoS channels (right) obtained with analog precoding.}
    \label{fig:analog}
\end{figure}

We evaluate the performance of analog precoding considering the achievable rate ratio, defined as $\Tilde{R}=R/C$, where $C=\log_2\left(1+\rho\lambda_{max}\right)$ is the single-stream channel capacity achieved with dominant eigenmode transmission, with $\lambda_{max}$ denoting the dominant eigenvalue of $\mathbf{T}=\mathbf{H}^H\mathbf{H}$.
In Fig.~\ref{fig:analog}, we report the performance of analog precoding based on our channel estimation strategy ``Refined''.
The baseline ``Perfect CSI'' refers to selecting the best analog precoder in the codebook for each channel realization.
Conversely, for the baseline ``No CSI'' we always use the precoder that is most frequently optimal within the training set.
We also compare our results with previous works employing \gls{lidar} data for beam selection \cite{kla19}-\cite{zec22}.
In the case of \gls{los} channels, precoding based on the refined channel achieves approximately the same rate as precoding based on $\mathbf{H}_{LoS}$.
In both \gls{los} and \gls{nlos} conditions, precoding based on the refined channel outperforms precoding based on $\mathbf{H}_{RT}$ and all previous works.

In Tab.~\ref{tab:complexity-CE-1}, we report the complexity of our \gls{cnn}, compared with \cite{kla19}-\cite{zec22}.
Our solution is less complex than \cite{kla19,dia19} but more complex than the solutions in \cite{mas21,zec22}.
This additional complexity is required by our solution since it explicitly estimates the \gls{mimo} channel matrix.
Conversely, in \cite{mas21,zec22}, the objective of the \gls{dl} architectures is merely to select the optimal beam within a predefined codebook.
Our more complex architecture brings three benefits.
First, by explicitly estimating the channel, we enable precoding strategies not limited to analog precoding, e.g., digital precoding.
Second, our strategy is agnostic about the considered precoding codebook.
Third, our strategy achieves higher performance than previous works using the same analog precoding codebook.

\begin{table}[t]
\centering
\caption{Complexity of precoding design solutions.}
\begin{tabular}{@{}ccc@{}}
\toprule
Model        & FLOPs & \# Params. \\
\midrule
CNN                & $152\times10^6$  & $297\times10^3$\\
\cite{kla19,dia19} & $1028\times10^6$ & $366\times10^3$\\
\cite{mas21}       & $3.50\times10^6$ & $6.82\times10^3$\\
\cite{zec22}       & $9.36\times10^6$ & $28.5\times10^3$\\
\bottomrule
\end{tabular}
\label{tab:complexity-CE-1}
\end{table}

\section{Conclusion}
\label{sec:conclusion}

We address the problems of blockage detection and precoder design in \gls{v2i} \gls{mimo} links, by completely avoiding communication overhead.
For blockage detection, we design a physics-based \gls{gnn} able to correctly classify the channel status with an accuracy of 95\%.
For precoder design, we propose a physics-based \gls{cnn} to refine a preliminary channel estimate obtained from \gls{lidar} data.
Digital precoding based on this refined channel estimate achieves 90\% of the capacity while analog precoding outperforms previous related works.

\bibliographystyle{IEEEtran}
\bibliography{IEEEabrv,main}

\begin{thebibliography}{10}
\providecommand{\url}[1]{#1}
\csname url@samestyle\endcsname
\providecommand{\newblock}{\relax}
\providecommand{\bibinfo}[2]{#2}
\providecommand{\BIBentrySTDinterwordspacing}{\spaceskip=0pt\relax}
\providecommand{\BIBentryALTinterwordstretchfactor}{4}
\providecommand{\BIBentryALTinterwordspacing}{\spaceskip=\fontdimen2\font plus
\BIBentryALTinterwordstretchfactor\fontdimen3\font minus
  \fontdimen4\font\relax}
\providecommand{\BIBforeignlanguage}[2]{{%
\expandafter\ifx\csname l@#1\endcsname\relax
\typeout{** WARNING: IEEEtran.bst: No hyphenation pattern has been}%
\typeout{** loaded for the language `#1'. Using the pattern for}%
\typeout{** the default language instead.}%
\else
\language=\csname l@#1\endcsname
\fi
#2}}
\providecommand{\BIBdecl}{\relax}
\BIBdecl

\bibitem{hur13}
S.~Hur, T.~Kim, D.~J. Love, J.~V. Krogmeier, T.~A. Thomas, and A.~Ghosh,
  ``{Millimeter Wave Beamforming for Wireless Backhaul and Access in Small Cell
  Networks},'' \emph{IEEE Trans. Commun.}, vol.~61, no.~10, pp. 4391--4403,
  2013.

\bibitem{hea16}
R.~W. Heath, N.~González-Prelcic, S.~Rangan, W.~Roh, and A.~M. Sayeed, ``{An
  Overview of Signal Processing Techniques for Millimeter Wave MIMO Systems},''
  \emph{IEEE J. Sel. Topics Signal Process.}, vol.~10, no.~3, pp. 436--453,
  2016.

\bibitem{gon17}
N.~Gonzalez-Prelcic, A.~Ali, V.~Va, and R.~W. Heath, ``{Millimeter-Wave
  Communication with Out-of-Band Information},'' \emph{IEEE Communications
  Magazine}, vol.~55, no.~12, pp. 140--146, 2017.

\bibitem{ali18}
A.~Ali, N.~González-Prelcic, and R.~W. Heath, ``{Millimeter Wave
  Beam-Selection Using Out-of-Band Spatial Information},'' \emph{IEEE Trans.
  Wireless Commun.}, vol.~17, no.~2, pp. 1038--1052, 2018.

\bibitem{va18}
V.~Va, J.~Choi, T.~Shimizu, G.~Bansal, and R.~W. Heath, ``{Inverse Multipath
  Fingerprinting for Millimeter Wave V2I Beam Alignment},'' \emph{IEEE Trans.
  Veh. Technol.}, vol.~67, no.~5, pp. 4042--4058, 2018.

\bibitem{alr20}
M.~Alrabeiah, A.~Hredzak, and A.~Alkhateeb, ``{Millimeter Wave Base Stations
  with Cameras: Vision-Aided Beam and Blockage Prediction},'' in \emph{2020
  IEEE 91st Vehicular Technology Conference (VTC2020-Spring)}, 2020, pp. 1--5.

\bibitem{cha21}
G.~Charan, M.~Alrabeiah, and A.~Alkhateeb, ``{Vision-Aided 6G Wireless
  Communications: Blockage Prediction and Proactive Handoff},'' \emph{IEEE
  Trans. Veh. Technol.}, vol.~70, no.~10, pp. 10\,193--10\,208, 2021.

\bibitem{kla19}
A.~Klautau, N.~González-Prelcic, and R.~W. Heath, ``{LIDAR Data for Deep
  Learning-Based mmWave Beam-Selection},'' \emph{IEEE Wireless Commun. Lett.},
  vol.~8, no.~3, pp. 909--912, 2019.

\bibitem{dia19}
M.~Dias, A.~Klautau, N.~González-Prelcic, and R.~W. Heath, ``{Position and
  LIDAR-Aided mmWave Beam Selection using Deep Learning},'' in \emph{2019 IEEE
  20th International Workshop on Signal Processing Advances in Wireless
  Communications (SPAWC)}, 2019, pp. 1--5.

\bibitem{mas21}
M.~B. Mashhadi, M.~Jankowski, T.-Y. Tung, S.~Kobus, and D.~Gündüz,
  ``{Federated mmWave Beam Selection Utilizing LIDAR Data},'' \emph{IEEE
  Wireless Commun. Lett.}, vol.~10, no.~10, pp. 2269--2273, 2021.

\bibitem{zec22}
M.~Zecchin, M.~B. Mashhadi, M.~Jankowski, D.~Gündüz, M.~Kountouris, and
  D.~Gesbert, ``{LIDAR and Position-Aided mmWave Beam Selection With Non-Local
  CNNs and Curriculum Training},'' \emph{IEEE Trans. Veh. Technol.}, vol.~71,
  no.~3, pp. 2979--2990, 2022.

\bibitem{lov03}
D.~Love and R.~Heath, ``{Equal gain transmission in multiple-input
  multiple-output wireless systems},'' \emph{IEEE Trans. Commun.}, vol.~51,
  no.~7, pp. 1102--1110, 2003.

\bibitem{kla18}
A.~Klautau, P.~Batista, N.~González-Prelcic, Y.~Wang, and R.~W. Heath, ``{5G
  MIMO Data for Machine Learning: Application to Beam-Selection Using Deep
  Learning},'' in \emph{2018 Information Theory and Applications Workshop
  (ITA)}, 2018, pp. 1--9.

\bibitem{dia11}
L.~{Di Angelo}, P.~{Di Stefano}, and L.~Giaccari, ``{A new mesh-growing
  algorithm for fast surface reconstruction},'' \emph{Computer-Aided Design},
  vol.~43, no.~6, pp. 639--650, 2011.

\end{thebibliography}
\end{document}